\documentclass[sigconf,screeen,table,xcdraw,anonymous=False]{acmart}

\acmConference[WSDM '21]{preparation for ACM Conference on Web Search and Data Mining}{March 8--12, 2021}{Jerusalem, Israel}
\acmBooktitle{preparation for ACM Conference on Web Search and Data Mining, March 8--12, 2021, Jerusalem, Israel}
\usepackage{times}

\usepackage{lineno}
\usepackage{booktabs}
\usepackage{multirow}
\usepackage{natbib}
\usepackage{enumitem}
\usepackage{balance}
\usepackage{flushend}

\usepackage{balance}

\usepackage{todonotes}
\presetkeys{todonotes}{inline}{}

\newcommand*{\origrightarrow}{}

\renewcommand*{\textrightarrow}{\fontfamily{cmr}\selectfont\origrightarrow}

\begin{document}

\title{Question Rewriting for Conversational Question Answering}


\author{Svitlana Vakulenko}
\authornote{This research was completed during the internship at Apple Inc.}
 \affiliation{\institution{University of Amsterdam}}
\email{s.vakulenko@uva.nl}

\author{Shayne Longpre}
    \affiliation{\institution{Apple Inc.}}
 \email{slongpre@apple.com}

\orcid{1234-5678-9012}
\author{Zhucheng Tu}
    \affiliation{\institution{Apple Inc.}}
\email{zhucheng_tu@apple.com}
 
\author{Raviteja Anantha}
    \affiliation{\institution{Apple Inc.}}
\email{raviteja_anantha@apple.com}






\begin{abstract}
Conversational question answering (QA) requires the ability to correctly interpret a question in the context of previous conversation turns.
We address the conversational QA task by decomposing it into question rewriting and question answering subtasks.
The question rewriting (QR) subtask is specifically designed to reformulate ambiguous questions, which depend on the conversational context, into unambiguous questions that can be correctly interpreted outside of the conversational context.
We introduce a conversational QA architecture that sets the new state of the art on the TREC CAsT 2019 passage retrieval dataset.
Moreover, we show that the same QR model improves QA performance on the QuAC dataset with respect to answer span extraction, which is the next step in QA after passage retrieval.
Our evaluation results indicate that the QR model we proposed achieves near human-level performance on both datasets and the gap in performance on the end-to-end conversational QA task is attributed mostly to the errors in QA.
\end{abstract}



\maketitle

\section{Introduction}
Extending question answering systems to a conversational setting is an important development towards a more natural human-computer interaction~\cite{DBLP:journals/ftir/GaoGL19}.
In this setting a user is able to ask several follow-up questions, which omit but reference information that was already introduced earlier in the same conversation, for example:
\smallskip
\begin{enumerate}[nosep]
\item[\textbf{(Q)}] - Where is Xi'an?
\item[\textbf{(A)}] - Shaanxi, China.
\item[\textbf{(Q)}] - What is \textbf{its} GDP?
\item[\textbf{(A)}] - 95 Billion USD.
\item[\textbf{(Q)}] - What is the share \textit{(of Xi'an)} in the \textit{(Shaanxi)} province GDP?
\item[\textbf{(A)}] - 41.8\% of Shaanxi's total GDP
\end{enumerate}
\smallskip
This example highlights two linguistic phenomena characteristic of a human dialogue, which include \textbf{anaphora} (words that explicitly reference previous conversation turns) and \textit{ellipsis} (words that can be omitted from the conversation)~\cite{thomas1979ellipsis}.
Therefore, conversational QA models require mechanisms capable of resolving contextual dependencies to correctly interpret such follow-up questions.
While existing co-reference resolution tools are designed to handle anaphoras, they do not provide any support for ellipsis~\cite{dalton2019trec}.
Previous research showed that QA models can be directly extended to incorporate conversation history but a considerable room for improvement still remains especially when scaling such models beyond a single input document~\cite{DBLP:conf/emnlp/ChoiHIYYCLZ18,DBLP:journals/tacl/ReddyCM19,DBLP:conf/sigir/Qu0CQCI20}.

In this paper we show how to extend existing state-of-the-art QA models with a QR component and demonstrate that the proposed approach improves the performance on the end-to-end conversational QA task.
This QR component is specifically designed to handle ambiguity of the follow-up questions by rewriting them such that they can be processed by existing QA models as stand-alone questions outside of the conversation context.
This setup also offers a wide range of practical advantages over the single end-to-end conversational QA model:
\begin{itemize}
    \item \textbf{Traceability}:
    QR component produces the question that the QA model is actually trying to answer, which allows to separate errors that stem from an incorrect question interpretation from errors in question answering.
    Our error analysis makes full use of this feature by investigating different sources of errors and their correlation.\\
    \item \textbf{Reuse}:
    QR allows to reduce the conversational QA task to the standard QA task and leverage already existing QA models and datasets.
    Any new non-conversational QA model can be immediately ported into a conversational setting using QR.
    A single QR model can be reused across several alternative QA architectures as we show in our experiments.\\
    \item \textbf{Modularity}:
   In practice, information is distributed across a network of heterogeneous nodes that do not share internal representations.
   A natural language question can provide an adequate communication protocol between these distributed components~\cite{DBLP:conf/naacl/RastogiGCM19}.
    Instead of sending the content of the whole conversation to a 3rd-party API, QR allows for formulating a concise question that contains only information relevant to the current information need, which also helps to determine which of the distributed systems should be used for answering the question.
\end{itemize}

Two dimensions on which QA tasks vary are: the type of data source used to retrieve the answer (e.g., a paragraph, a document collection, or a knowledge graph); and the expected answer type (a text span, a ranked list of passages, or an entity).
In this paper, we experiment with two variants of the QA task: \textit{retrieval QA}, the task of finding an answer to a given natural-language question as a ranked list of relevant passages given a document collection; and \textit{extractive QA}, the task of finding an answer to a given natural-language question as a text span within a given passage.
Though the two QA tasks are complementary to each other, in this paper we focus on the QR task and its ability to enable different types of QA models within a conversational setting.
We experiment with both retrieval and extractive QA models to examine the effect of the QR component on the end-to-end QA performance.
%
The contributions of this work are three-fold:
\begin{enumerate}
\item We introduce a novel approach for the conversational QA task based on question rewriting that sets the new state-of-the-art results on the TREC CAsT dataset for passage retrieval.
\item We show that the same question rewriting model (trained on the same dataset) boosts the performance of the state-of-the-art architecture for answer extraction on the QuAC dataset.
\item We systematically evaluate the proposed approach in both retrieval and extractive settings, and report the results of our error analysis.
\end{enumerate}

\section{Related Work}
\label{sec:rel_work}

\textit{\textbf{Conversational QA}} is an extension of the standard QA task that introduces contextual dependencies between the input question and the previous dialogue turns.
Several datasets were recently proposed extending different QA tasks to a conversational setting including extractive~\cite{DBLP:journals/tacl/ReddyCM19,DBLP:conf/emnlp/ChoiHIYYCLZ18}, retrieval~\cite{dalton2019trec} and knowledge graph QA~\cite{DBLP:conf/nips/GuoTDZY18,christmann2019look}.
One common approach to conversational QA is to extend the input of a QA model by appending previous conversation turns~\cite{DBLP:conf/cikm/QuYQZCCI19,DBLP:journals/corr/abs-1909-10772,christmann2019look}.
Such approach, however falls short in case of retrieval QA, which requires a concise query as input to the candidate selection step, such as BM25~\cite{DBLP:journals/corr/abs-1904-08375}.
Results of the recent TREC CAsT track demonstrated that co-reference models are also not sufficient to resolve the missing context in the follow-up questions~\cite{dalton2019trec}.
A considerable gap between the performance of automated rewriting approaches and manual human annotations call for new architectures that are capable of retrieving relevant answers from large text collections using conversational context.


\textit{\textbf{Query rewriting}} is a technique that was successfully applied in a variety of tasks, including data integration, query optimization and query expansion in sponsored search~\cite{DBLP:conf/sigmod/PapakonstantinouV99,DBLP:conf/sigmod/RizviMSR04,DBLP:conf/sigir/GrbovicDRSB15}.
More recently rewriting conditioned on the conversation history was shown to increase effectiveness of chitchat and task-oriented dialogue models~\cite{DBLP:conf/acl/SuSZSHNZ19,DBLP:conf/naacl/RastogiGCM19}.
QR in the context of the conversational QA task was first introduced by \citeauthor{elgohary2019can}~\cite{elgohary2019can}, who released the CANARD dataset that contains human rewrites of questions from QuAC conversational reading comprehension dataset.
Their evaluation, however, was limited to analysing rewriting quality without assessing impact from rewriting on the end-to-end conversational QA task.
Our study was designed to bridge this gap and extend evaluation of question rewriting for conversational QA to the passage retrieval task as well.
The field of conversational QA is advancing rapidly due to the challenges organised by the community, such as the TREC CAsT challenge~\cite{dalton2019trec}.
Several recent studies that are concurrent to our work proposed alternative approaches for the conversational QA task and evaluated them on the TREC CAsT dataset.
\citet{DBLP:conf/sigir/MeleMN0TF20} introduced an unsupervised approach that relies on a set of heuristics using part-of-speech tags, dependency parses and co-reference resolution to rewrite questions.
\citeauthor{DBLP:conf/sigir/VoskaridesLRKR20}~\cite{DBLP:conf/sigir/VoskaridesLRKR20} train a binary classification model on CANARD dataset that learns to pick the terms from the conversation history for query expansion.
\citeauthor{DBLP:conf/sigir/YuLYXBG020}~\cite{DBLP:conf/sigir/YuLYXBG020} train a sequence generation model using a set of rules applied to the MS MARCO search sessions.
We demonstrate that our approach does not only outperform all the results reported on the TREC CAsT dataset previously but also boosts the performance on the answer extraction task as demonstrated on the QuAC dataset.


\section{Question Rewriting Task}
\label{sec:task}

We assume the standard setting of a conversational (sequential) QA task, in which a user asks a sequence of questions and the system is to provide an answer after each of the questions.
Every follow-up question may be ambiguous, i.e., its interpretation depends on the previous conversation turns (example: ``What is its GDP?'').
The task of question rewriting is to translate every ambiguous follow-up question into a semantically equivalent but unambigous question (for this example: ``What is the GDP of Xi'an?'').
Then, every question is first processed by the QR component before passing it to the QA component.

\begin{figure*}[h]
  \centering
  \includegraphics[width=0.9\linewidth]{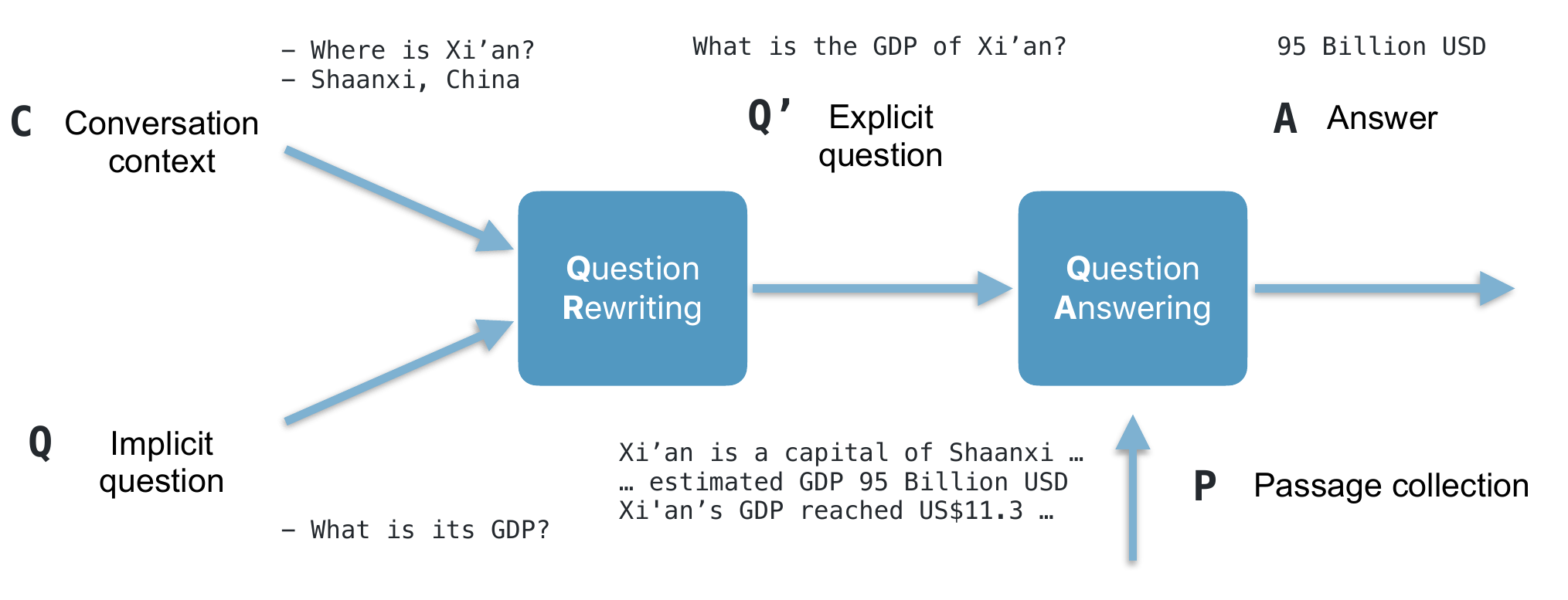}
  \caption{Our approach for end-to-end conversational QA relies on the question rewriting component to handle conversation context and produce an explicit question that can be fed to standard, non-conversational QA components.}
  \label{fig:architecture}
\end{figure*}

More formally, given a conversation context $C$ and a potentially implicit question $Q$, a question which may require the conversation context $C$ to be fully interpretable, the task of a question rewriting (QR) model is to generate an explicit question $Q'$ which is equivalent to $Q$ under conversation context $C$ and has the same correct answer $A$.
See Figure~\ref{fig:architecture} for an illustrative example that shows how QR can be combined with any non-conversational QA model.

\section{Approach}
\label{sec:approach}


In this section, we describe the model architectures of our QR and QA components.
We use two distinct QA architectures to show that the same QR model can be successfully applied across several QA models irrespective of their implementation.
One of the QA models is designed for the task of passage retrieval and the other one for the task of answer extraction from a passage (reading comprehensions).
We employ state-of-the-art QA architectures that were already successfully evaluated for these tasks in non-conversational settings and show how QR allows to port existing QA models into a conversational setting.


\subsection{Question Rewriting Model}
\label{sec:ag_eqs}

We use a model for question rewriting, which employs a unidirectional Transformer decoder~\cite{radford2019language} for both encoding the input sequence and decoding the output sequence.
The input to the model is the question with previous conversation turns (we use 5 previous turns in our experiments) turned into token sequences separated with a special $[SEP]$ token.

The training objective is to predict the output tokens provided in the ground truth question rewrites produced by human annotators.
The model is trained via teacher forcing approach, which is a standard technique for training language generation models, to predict every next token in the output sequence given all the preceding tokens.
The loss is calculated via negative log-likelihood (cross-entropy) between the output distribution $D' \in \mathbb{R}^{|V|}$ over the vocabulary $V$, and the one-hot vector $y_{Q'} \in \mathbb{R}^{|V|}$ for the correct token from the ground truth: $loss = - y_{Q'}\log D'$.

\begin{figure}[h]
  \centering
  \includegraphics[width=\linewidth]{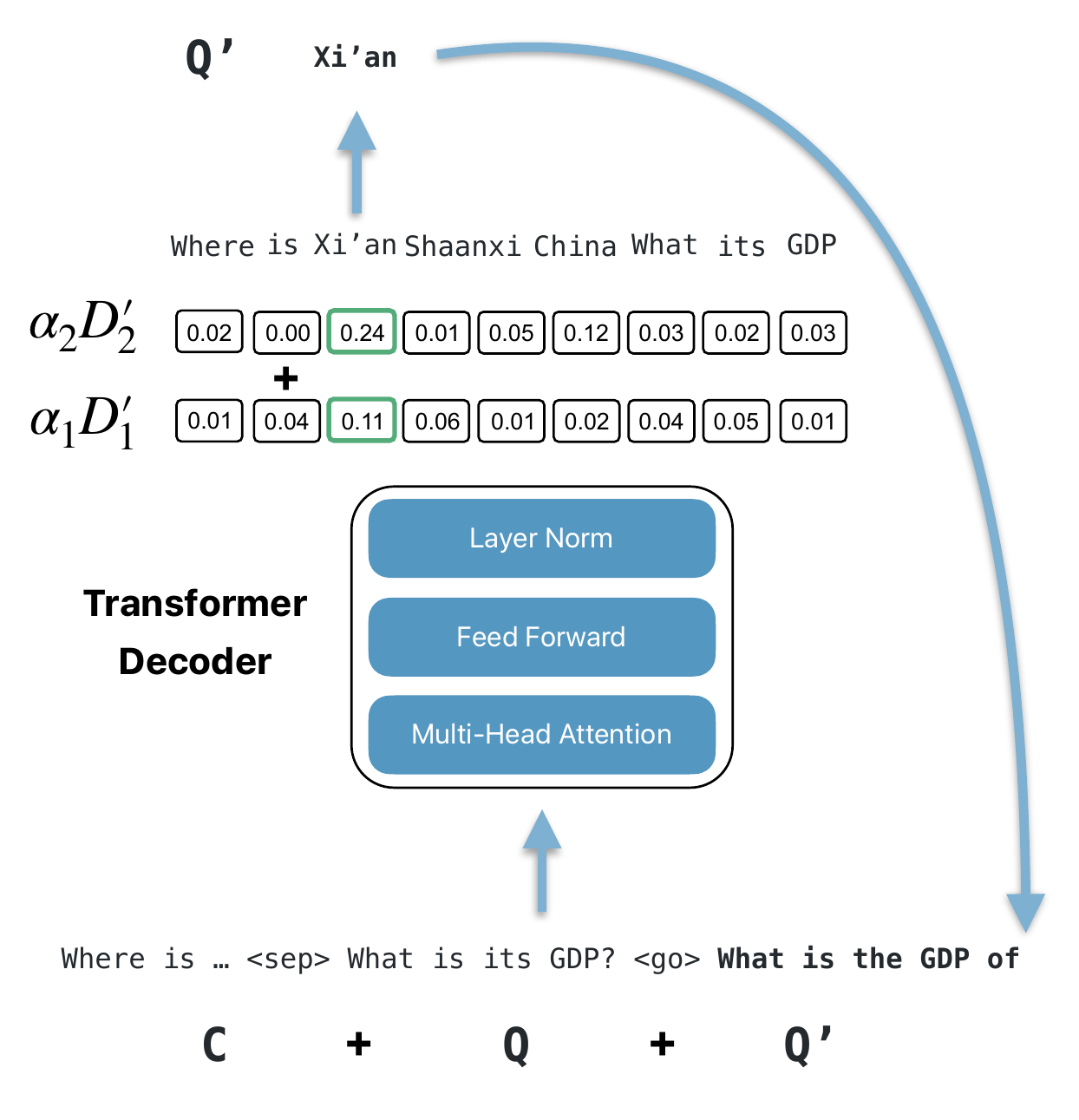}
  \caption{The question rewriting component uses the Transformer Decoder architecture, to recursively generate the tokens of an "explicit" question. At inference time, the generated output is appended to the input sequence for the next timestep in the sequence.}
  \label{fig:qr}
\end{figure}

At training time the output sequence is shifted by one token and is used as input to predict all next tokens of the output sequence at once.
At inference time, the model uses maximum likelihood estimation to select the next token from the final distribution $D'$ (greedy decoding), as shown in Figure~\ref{fig:qr}.

We further increase capacity of our generative model by learning to combine several individual distributions ($D'_{1}$ and $D'_{2}$ in Figure~\ref{fig:qr}).
The final distribution $D'$ is then produced as a weighted sum of the intermediate distributions: $D' = \sum_{i=0}^{m} \alpha_{i}D'_{i}$ ($m = 2$ in our experiments).
To produce $D'_{i} \in \mathbb{R}^{|V|}$ we pass the last hidden state of the Transformer Decoder $h \in \mathbb{R}^{d}$ through a separate linear layer for each intermediary distribution: $D'_{i} = W^{H}_{i}h + b^H$, where $W^{H}_{i}$ is the weight matrix and $b^H$ is the bias.
For the weighting coefficients $\alpha_{i}$ we use the matrix of input embeddings $X \in \mathbb{R}^{n \times d}$, where $n$ is the maximum sequence length and $d$ is the embedding dimension, and the output of the first attention head of the Transformer Decoder $G \in \mathbb{R}^{n \times d}$ put through a layer normalization function: $\alpha_{i} = W^{G}_{i} norm(G) + W^{X}_{i}X + b^{\alpha}_{i}$, where all $W$ are the weight matrices and $b^{\alpha}_{i}$ is the bias.

\subsection{Retrieval QA Model}
In the retrieval QA settings, the task is to produce a ranked list of text passages from a collection, ordered by their relevance to a given a natural language question~\cite{DBLP:conf/nips/NguyenRSGTMD16,dietz2018trec}.
We employ the state-of-the-art approach to retrieval QA, which consists of two phases: candidate selection and passage re-ranking.
This architecture holds the state-of-the-art for the passage retrieval task on the MS MARCO dataset according to the recent experiments conducted by \citeauthor{xiong2020approximate}~\cite{xiong2020approximate}.

In the first phase, a traditional retrieval algorithm (BM25) is used to quickly sift through the indexed collection retrieving top-$k$ passages ranked by relevance to the input question $Q'$.
In the second phase, a more computationally-expensive model is used to re-rank all question-answer candidate pairs formed using the previously retrieved set of $k$ passages.

\begin{figure}[h]
  \centering
  \includegraphics[width=\linewidth]{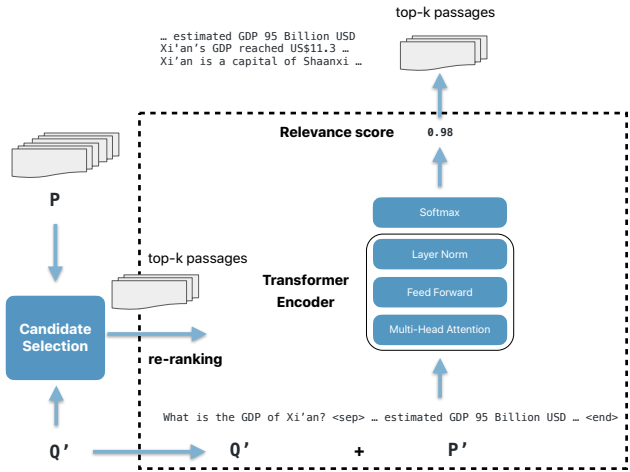}
  \caption{Retrieval QA component includes two sequential phases: candidate selection (BM25) followed by passage re-ranking (Transformer Encoder).}
    \label{fig:rqa_architecture}
\end{figure}

For re-ranking, we use a binary classification model that predicts whether the passage answers a question, i.e., the output of the model is the relevance score in the interval $[0, 1]$.
The input to the re-ranking model is the concatenated question and passage with a separation token in between (see Figure~\ref{fig:rqa_architecture} for the model overview).
The model is initialized with weights learned from unsupervised pre-training on the language modeling (masked token prediction) task (BERT)~\cite{devlin2018bert}.
During fine-tuning, the training objective is to reduce cross-entropy loss, using relevant passages and non-relevant passages from the top-$k$ candidate passages.

\subsection{Extractive QA Model}
\label{section:extractive_qa_approach}
The task of extractive QA is given a natural language question and a single passage find an answer as a contiguous text span within the given passage~\cite{DBLP:conf/emnlp/RajpurkarZLL16}.
Our model for extractive QA consists of a Transformer-based bidirectional encoder (BERT)~\cite{devlin2018bert} and an output layer predicting the answer span.
This type of model architecture corresponds to the current state-of-the-art setup for several reading comprehension benchmarks~\citep{liu2019roberta, lan2019albert}.

The input to the model is the sequence of tokens formed by concatenating a question and a passage separated with a special $[SEP]$ token.
The encoder layers are initialized with the weights of a Transformer model pre-trained on an unsupervised task (masked token prediction).
The output of the Transformer encoder is a hidden vector $T_i$ for each token $i$ of the input sequence.

\begin{figure}[h]
  \centering
  \includegraphics[width=\linewidth]{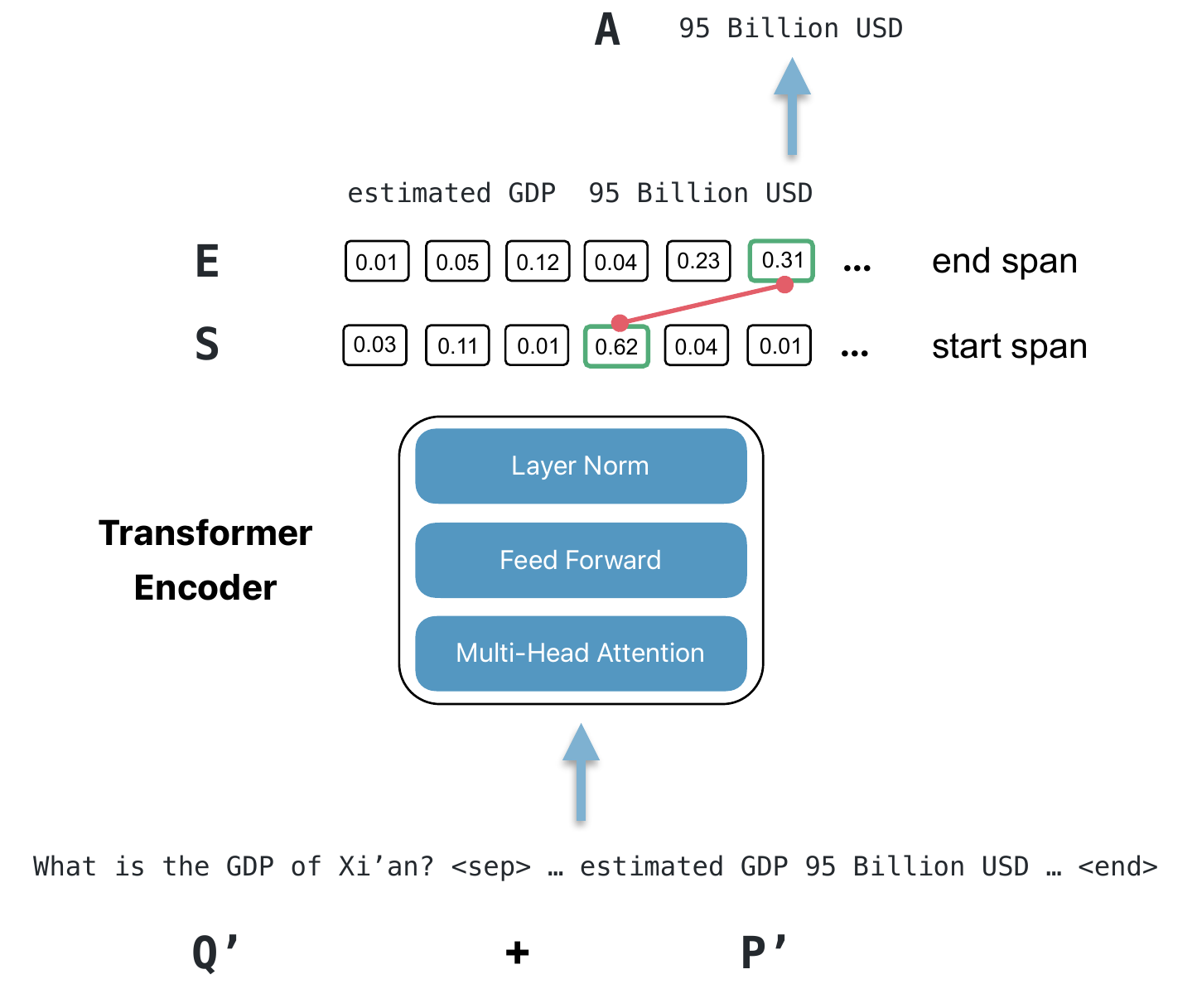}
  \caption{Extractive QA component predicts a span of text in the paragraph \textbf{P'}, given an input sequence with the question \textbf{Q'} and passage \textbf{P'}.}
  \label{fig:eqa_architecture}
\end{figure}

For fine-tuning the model on the extractive QA task, we add weight matrices $W^s$, $W^e$ and biases $b^s$, $b^e$ that produce two probability distributions over all the tokens of the given passage separately for the start ($S$) and end position ($E$) of the answer span.
For each token $i$ the output of the Transformer encoder $T_i$ is passed through a linear layer, followed by a softmax normalizing the output logits over all the tokens into probabilities:
\begin{equation}
    S_i = \frac{e^{W^{s}{\cdot}T_i+b^s}}{\sum_{j=1}^{n} e^{W^{s}{\cdot}T_j+b^s}}\;\;\;\;\;\;E_i = \frac{e^{W_{e}{\cdot}T_i+b_e}}{\sum_{j=1}^{n} e^{W^{e}{\cdot}T_j+b^e}}\label{eq:1}
\end{equation}
The model is then trained to minimize cross-entropy between the predicted start/end positions ($S_i$ and $E_i$) and the correct ones from the ground truth ($y_S$ and $y_E$ are one-hot vectors indicating the correct start and end tokens of the answer span):
\begin{equation}
    loss = - \sum_{i=1}^{n} y_S\log S_i - \sum_{i=1}^{n} y_E\log E_i
\end{equation}
At inference time all possible answer spans from position $i$ to position $j$, where $j \geq i$, are scored by the sum of end and start positions' probabilities: $S_i + E_j$.
The output of the model is the maximum scoring span (see Figure~\ref{fig:eqa_architecture} for the model overview). 

21\% of the CANARD (QuAC) examples are Not Answerable (NA) by the provided passage. 
To enable our model to make No Answer predictions we prepend a special $[CLS]$ token to the beginning of the input sequence.
For all No Answer samples we set both the ground truth start and end span positions to this token's position (0).
Likewise, at inference time, predicting this special token is equivalent to a No Answer prediction for the given example.

\section{Experimental Setup}
\label{sec:experiments}
To evaluate the question rewriting approach we perform a range of experiments with several existing state-of-the-art QA models.
In the following subsections we describe the datasets used for training and evaluation, the set of metrics for each of the components, our baselines and details of the implementation.
Our experimental setup is designed to evaluate gains in performance, reuse and traceability from introducing the QR component:
\paragraph{\textbf{RQ1}: How does the proposed approach perform against competitive systems (performance)?}
\paragraph{\textbf{RQ2}: How does non-conversational pre-training benefit the models with and without QR (reuse)?}
\paragraph{\textbf{RQ3}: What is the proportion of errors contributed by each of the components (traceability)?}

\subsection{Datasets}
We chose two conversational QA datasets for the evaluation of our approach: (1) CANARD, derived from Question Answering in Context (QuAC) for extractive conversational QA \cite{DBLP:conf/emnlp/ChoiHIYYCLZ18}, and (2) TREC CAsT for retrieval conversational QA~\cite{dalton2019trec}.
See Table~\ref{tab:datasets} for the overview of the datasets.
Since TREC CAsT is relatively small we used only CANARD for training QR.
The same QR model trained on CANARD is evaluated on both CANARD and TREC CAsT.


CANARD \cite{elgohary2019can} contains sequences of questions with answers as text spans in a given Wikipedia article.
It was built upon the QuAC dataset~\cite{DBLP:conf/emnlp/ChoiHIYYCLZ18} by employing human annotators to rewrite original questions from QuAC dialogues into explicit questions.
CANARD consists of 40.5k pairs of question rewrites that can be matched to the original answers in QuAC.
We use CANARD splits for training and evaluation.
We use the question rewrites provided in CANARD and articles with correct answer spans from QuAC.
In our experiments, we refer to this joint dataset as CANARD for brevity.

To further boost performance of the extractive QA model we reuse the approach from \citeauthor{fisch2019mrqa}~\cite{fisch2019mrqa} and pre-train the model on MultiQA dataset, which contains $75k$ QA pairs from six standard QA benchmarks, and then fine-tune it on CANARD to adapt for the domain shift.
Note that MultiQA is a non-conversational QA dataset.

TREC CAsT contains sequences of questions with answers to be retrieved as text passages from a large collection of passages: MS MARCO~\cite{DBLP:conf/nips/NguyenRSGTMD16} with 8.6M passages + TREC CAR \cite{dietz2018trec} with 29.8M passages.
The official test set that we used for evaluation contains relevance judgements for 173 questions across 20 dialogues (topics).
We use the model from \citeauthor{nogueira2019passage}~\cite{nogueira2019passage} for retrieval QA, which was tuned on a sample from MS MARCO with 12.8M query-passage pairs and 399k unique queries.



\begin{table}[]
  \caption{Datasets used for training and evaluation (with the number of questions).}
  \label{tab:datasets}
    \begin{tabular}{cccc}
        \toprule
        & Question &  Retrieval  & Extractive \\
        \multirow{-2}{*}{Tasks} & Rewriting &  QA  &  QA  \\
            \midrule
             & & MS MARCO & +MultiQA (75k)\\
           \multirow{-2}{*}{Train} & \multirow{-2}{*}{CANARD (35k)} &  (399k) & CANARD (35k) \\ 
           \midrule
            & TREC CAsT (173)& TREC CAsT & \\ 
        \multirow{-2}{*}{Test} & CANARD (5.5k) & (173) & \multirow{-2}{*}{CANARD (5.5k)} \\ 
        \bottomrule
    \end{tabular}
\end{table}

\subsection{Metrics}
Our evaluation setup corresponds to the one used in TREC CAsT, which makes our experiments directly comparable to the official TREC CAsT results.
Mean average precision (\textit{MAP}), mean reciprocal rank (\textit{MRR}), normalized discounted cumulative gain (\textit{NDCG@3}) and precision on the top-passage (\textit{P@1}) evaluate quality of passage retrieval.
Top-1000 documents are considered per query with a relevance judgement value cut-off level of 2 (the range of relevance grades in TREC CAsT is from 0 to 4).

We use \textit{F1} and Exact Match (\textit{EM}) for extractive QA, which measure word token overlap between the predicted answer span and the ground truth.
We also report accuracy for questions without answers in the given passage (\textit{NA Acc}).

Our analysis showed that \textit{ROUGE} recall calculated for unigrams (ROUGE-1 recall) correlates with the human judgement of the question rewriting performance (Pearson 0.69), which we adopt for our experiments as well.
ROUGE~\cite{lin2004rouge} is a standard metric of lexical overlap, which is often used in text summarization and other text generation tasks.
We also calculate question similarity scores with the Universal Sentence Encoder (\textit{USE}) model~\cite{DBLP:conf/emnlp/CerYKHLJCGYTSK18} (Pearson 0.71).


\subsection{QR Baselines}
\label{sec:qr_models}

The baselines were designed to challenge the need for a separate QR component by incorporating previous turns as direct input to custom QA components.
Manual rewrites by human annotators provide the upper-bound performance for a QR approach and allows for an ablation study of the down-stream QA components.

\paragraph{\textbf{Original.}} Original questions from the conversational QA datasets without any question rewriting. 

\paragraph{\textbf{Original + $k$-DT.}}
Our baseline approach for extractive QA prepends the previous $k$ questions to the original question to compensate for the missing context. The questions are separated with a special token and used as input to the Transformer model. We report the results for $k=\{1,2,3\}$.

\paragraph{\textbf{Original + $k$-DT*.}}
Since in the first candidate selection phase we use BM25 retrieval function which operates on a bag-of-words representation, we modify the baseline approach for retrieval QA as follows.
We select keywords from $k$ prior conversation turns (not including current turn) based on their inverse document frequency (IDF) scores and append them to the original question of the current turn.
We use the keyword-augmented query as the search query for \textbf{Anserini}~\cite{Yang:2017:AEU:3077136.3080721}, a Lucene toolkit for replicable information retrieval research, and if we use BERT re-ranking we concatenate the keyword-augmented query with the passages retrieved from the keyword-augmented query.
We use the keywords with IDF scores above the threshold of 0.0001, which was selected based on a 1 million document sample of the MS MARCO corpus.

\paragraph{\textbf{Human.}} To provide an upper bound, we evaluate all our models on the question rewrites manually produced by human annotators.

\subsection{QR Models}

In addition to the baselines described above, we chose several alternative models for question rewriting of the conversational context: (1) co-reference resolution as in the TREC CAsT challenge; (2) PointerGenerator used for question rewriting on CANARD by \citet{elgohary2019can} but not previously evaluated on the end-to-end conversational QA task; (3) CopyTransformer extension of the PointerGenerator model that replaces the bi-LSTM encoder-decoder architecture with a Transformer Decoder model.
All models, except for co-reference, were trained on the train split of the CANARD dataset.
Question rewrites are generated turn by turn for each dialogue recursively using already generated rewrites as previous turns. 
This is the same setup as in the TREC CAsT evaluation.

\paragraph{\textbf{Co-reference.}} Anaphoric expressions in original questions are replaced with their antecedents from the previous dialogue turns.
Co-reference dependencies are detected using a publicly available neural co-reference resolution model that was trained on OntoNotes~\cite{DBLP:conf/naacl/LeeHZ18}.\footnote{\url{https://github.com/kentonl/e2e-coref}}

\paragraph{\textbf{PointerGenerator.}} A sequence-to-sequence model for text generation with bi-LSTM encoder and a pointer-generator decoder~\cite{elgohary2019can}.

\paragraph{\textbf{CopyTransformer.}} The Transformer decoder, which, similar to pointer-generator model, uses one of the attention heads as a pointer~\cite{DBLP:conf/emnlp/GehrmannDR18}.
The model is initialized with the weights of a pre-trained GPT2 model~\cite{radford2019language,Wolf2019HuggingFacesTS} (Medium-sized GPT-2 English model: 24-layer, 1024-hidden, 16-heads, 345M parameters) and then fine-tuned on the question rewriting task.

\paragraph{\textbf{Transformer++.}} The Transformer-based model described in Section~\ref{sec:ag_eqs}. Transformer++ is initialized with the weights of the pre-trained GPT2 model, same as in CopyTransformer.

\subsection{QA Models}

Our retrieval QA approach is implemented as proposed in~\cite{nogueira2019passage} using Anserini for the candidate selection phase with BM25 (top-1000 passages) and $BERT_{LARGE}$ for the passage re-ranking phase (Anserini + BERT).
Both components were fine-tuned only on the MS MARCO dataset ($k_1=0.82, b=0.68$).\footnote{\url{https://github.com/nyu-dl/dl4marco-bert}}

We train several models for extractive QA on different variants of the training set based on the CANARD training set~\cite{elgohary2019can}.
All models are first initialized with the weights of the $BERT_{LARGE}$ model pre-trained using the whole word masking~\cite{devlin2018bert}.

\paragraph{\textbf{CANARD-O.}} The baseline models were trained using original (implicit) questions of the CANARD training set with a dialogue context of varying length (Original and Original + $k$-DT).
The models are trained separately for each $k=\{0,1,2,3\}$, where $k=0$ corresponds to the model trained only on the original questions without any previous dialogue turns.

\paragraph{\textbf{CANARD-H.}} To accommodate input of the question rewriting models, we train a QA model that takes human rewritten question from the CANARD dataset as input without any additional conversation context, i.e., as in the standard QA task.

\paragraph{\textbf{MultiQA $\rightarrow$ CANARD-H.}} Since the setup with rewritten questions does not differ from the standard QA task, we experiment with pretraining the extractive QA model on the MultiQA dataset with explicit questions~\cite{fisch2019mrqa}, using parameter choices introduced by ~\citet{longpre2019exploration}.
We further fine-tune this model on the target CANARD dataset to better adopt it for the domain shift in CANARD QA samples (see Figure~\ref{fig:canard-partial_data}).

\section{Results}
\label{sec:results}




\paragraph{\textbf{RQ1}: How does the proposed approach perform against competitive systems (performance)?}

Our approach, using question rewriting for conversational QA, consistently outperforms the baselines that use previous dialogue turns, in both retrieval and extractive QA tasks (see Tables~\ref{table:trec}-\ref{table:quac}).
Moreover, it shows considerable improvement over the latest results reported on the TREC CAsT dataset (see Table~\ref{table:trec_soa}).

\begin{table}[t]
	\centering
	\caption{Evaluation results of the QR models. *Human performance is measured as the difference between two independent annotators' rewritten questions, averaged over 100 examples. This provides an estimate of the upper bound.}
	\label{table:qr-results}
			\begin{tabular}{llccc}
				\toprule
				Test Set & Question & ROUGE & USE & EM\\
				\midrule
				CANARD & Original & 0.51 & 0.73 & 0.12 \\
				 & Co-reference & 0.68 & 0.83 & 0.48 \\
				 & PointerGenerator & 0.75 & 0.83 & 0.22 \\
				 & CopyTransformer & 0.78 & 0.87 & 0.56 \\
				 & Transformer++ & \textbf{0.81} & \textbf{0.89} & \textbf{0.63} \\
				 \midrule
				 & Human* & 0.84 & 0.90 & 0.33 \\
				 \midrule
				 TREC CAsT & Original & 0.67 & 0.80 & 0.28\\
				   & Co-reference & 0.71  & 0.80 & 0.13\\
				 & PointerGenerator & 0.71 & 0.82 & 0.17\\
				 & CopyTransformer & 0.82 & 0.90 & 0.49 \\
				 & Transformer++ & \textbf{0.90} & \textbf{0.94} & \textbf{0.58}\\
				 \midrule
				 & Human* & 1.00 & 1.00 & 1.00\\
				\bottomrule
			\end{tabular}
     \vspace{2mm}
\end{table}

\begin{figure}[h]
  \centering
  \includegraphics[width=\linewidth]{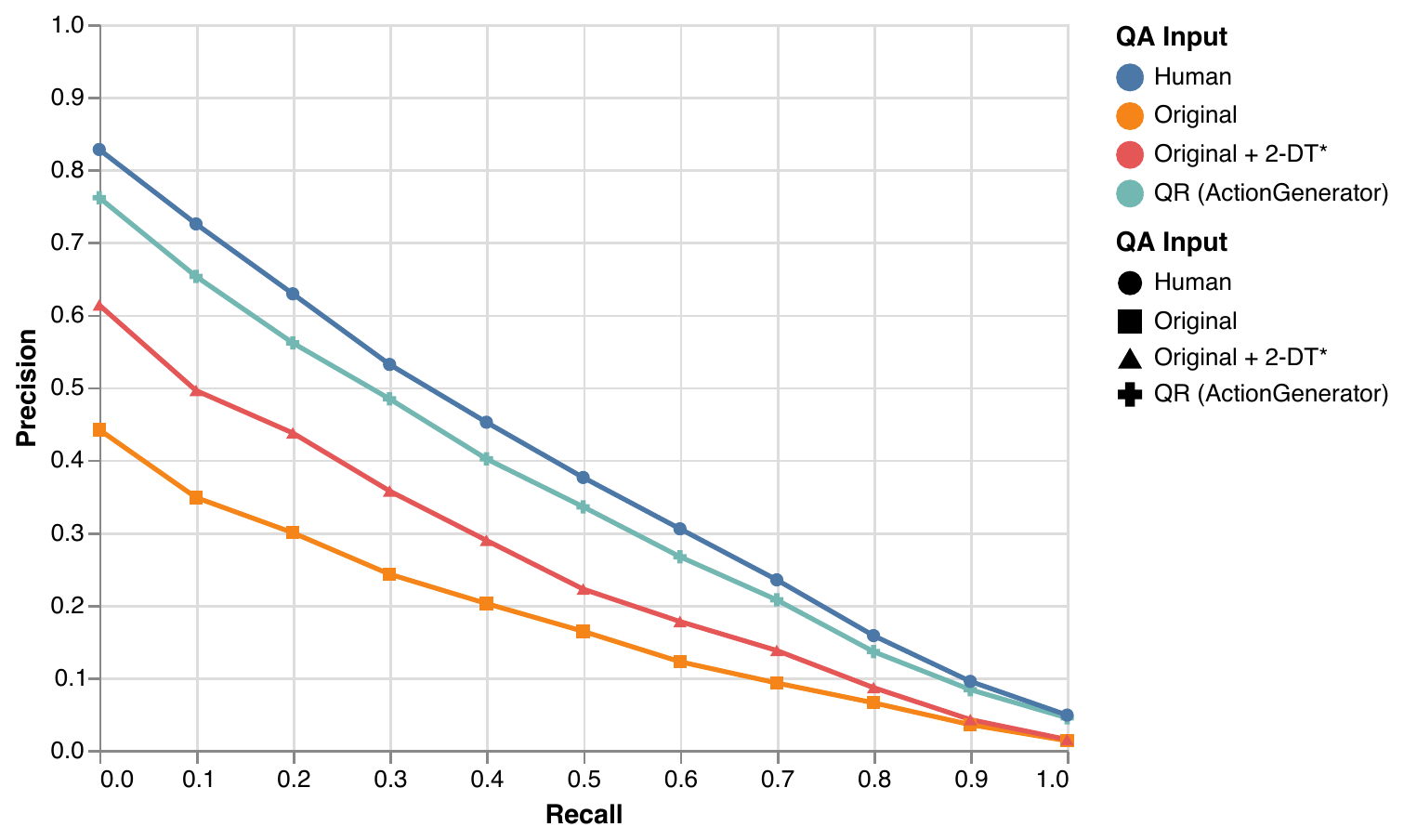}
  \caption{Precision-recall curve illustrating model performance on the TREC CAsT test set for Anserini + BERT.}
  \label{fig:trec_precision_recall}
\end{figure}

The precision-recall trade-off curve in Figure~\ref{fig:trec_precision_recall} shows that question rewriting performance is close to the performance achieved by manually rewriting implicit questions.
Our results also indicate that the QR performance metric is able to correctly predict the model that performs consistently better across both QA tasks (see Table~\ref{table:qr-results}).

Passage re-ranking with BERT always improves ranking results (almost a two-fold increase in MAP, see Table~\ref{table:trec}).
Keyword-based baselines (Original + $k$-DT*) prove to be very strong outperforming both Co-reference and PointerGenerator models on all three performance metrics.
Both MRR and NDCG@3 are increasing with the number of turns used for sampling keywords, while MAP is slightly decreasing, which indicates that it brings more relevant results at the very top of the rank but non-relevant results also receive higher scores.
In contrast, the baseline results for Anserini + BERT model indicate that the re-ranking performance for all metrics decreases if the keywords from more than 2 previous turns are added to the original question.

Similarly for extractive QA, the model incorporating previous turns proved to be a very strong baseline (see Table~\ref{table:quac}).
The performance results also suggest that all models do not discriminate well the passages that do not have an answer to the question (71\% accuracy on the human rewrites).
We notice that the baseline models, which were trained with the previous conversation turns, tend to adopt a conservative strategy by answering more questions as ``Unanswerable'' (NA).
Controlling for this effect, we show that our QR model gains a higher performance level by actually answering questions that have answers.

\begin{table}[t]
	\centering
    \caption{Retrieval QA results on the TREC CAsT test set.}
    	\label{table:trec}
			\begin{tabular}{llccc}
				\toprule
				QA Input & QA Model & MAP & MRR & NDCG@3 \\
				\midrule
	            Original & Anserini & 0.089 & 0.245 & 0.131 \\
				Original + 1-DT* &  & 0.133 & 0.343 & 0.199 \\
				Original + 2-DT* &  & 0.130 & 0.374 & 0.213 \\
				Original + 3-DT* &  & 0.127 & 0.396 & 0.223 \\
				{Co-reference} &  & 0.109 & 0.298 & 0.172 \\
				{PointerGenerator} &  & 0.100 & 0.273 & 0.159 \\
				{CopyTransformer} &  & 0.148 & 0.375 & 0.213 \\
				Transformer++ &  & 0.190 & 0.441 & 0.265 \\
				{Human} &  & 0.218 & 0.500 & 0.315 \\
				\midrule
			    Original & Anserini & 0.172 & 0.403 & 0.265 \\
				Original + 1-DT* & +BERT & 0.230 & 0.535 & 0.378 \\
				Original + 2-DT* & & 0.245 & 0.576 & 0.404 \\
				Original + 3-DT* &  & 0.238 & 0.575 & 0.401 \\
				{Co-reference} &  & 0.201 & 0.473 & 0.316 \\
				{PointerGenerator} &  & 0.183 & 0.451 & 0.298 \\
				CopyTransformer &  & 0.284 & 0.628 & 0.440 \\
				Transformer++ &  & \textbf{0.341} & \textbf{0.716} & \textbf{0.529} \\
				{Human} &  & 0.405 & 0.879 & 0.589 \\
				\bottomrule
			\end{tabular}
\end{table}

\begin{table}[t]
	\centering
    \caption{Comparison with the state-of-the-art results reported on the TREC CAsT test set.}
    	\label{table:trec_soa}
			\begin{tabular}{lc}
				\toprule
				Approach & NDCG@3 \\
				\midrule
				\citeauthor{DBLP:conf/sigir/MeleMN0TF20}~\cite{DBLP:conf/sigir/MeleMN0TF20} & 0.397  \\
				\citeauthor{DBLP:conf/sigir/VoskaridesLRKR20}~\cite{DBLP:conf/sigir/VoskaridesLRKR20} & 0.476  \\
				\citeauthor{DBLP:conf/sigir/YuLYXBG020}~\cite{DBLP:conf/sigir/YuLYXBG020} & 0.492 \\
				Ours & \textbf{0.529} \\
				\bottomrule
			\end{tabular}
\end{table}



\paragraph{\textbf{RQ2}: How does non-conversational pre-training benefit the models with and without QR (reuse)?}

\begin{figure}[h]
  \centering
    \includegraphics[width=\linewidth]{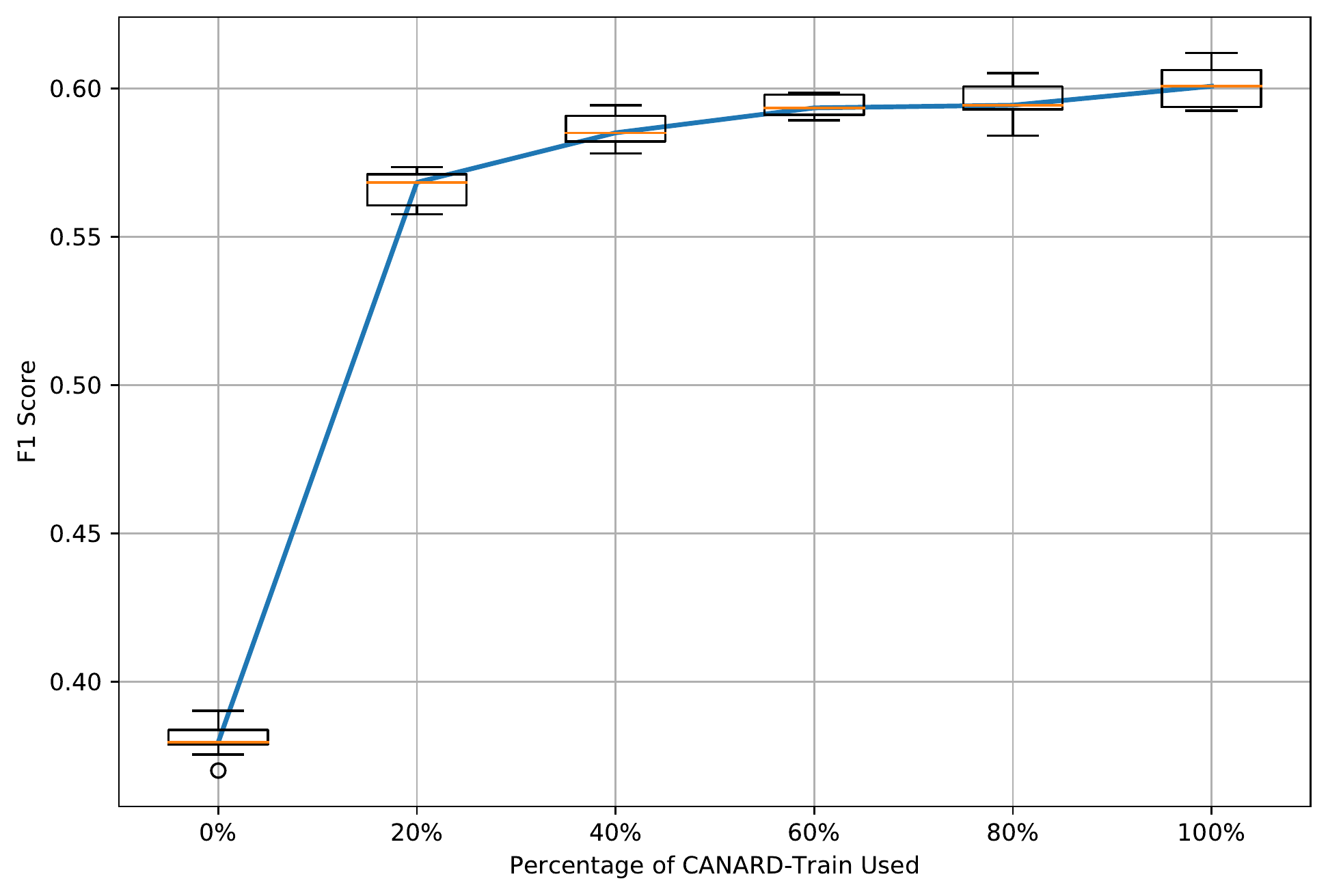}
    \caption{Effect from fine-tuning the MultiQA model on a portion of the target CANARD-H dataset due to the domain shift between the datasets.}
    \label{fig:canard-partial_data}
\end{figure}

We observe that pre-training on MultiQA improves performance of all extractive QA model.
However, it is much more prominent for the systems using question rewrites (7\% increase in EM and 6\% in F1 when using human-rewritten questions).
The models do not require any additional fine-tuning, when using QR, since the type of input to the QA model remains the same (non-conversational).
While for Original + $k$-DT models fine-tuning is required also to adopt the model for the new type of input data (conversational).
Note that, in this case, we had to fine-tune all models but for another reason.
The style of questions in CANARD is rather different from other QA datasets in MultiQA.
Figure~\ref{fig:canard-partial_data} demonstrates that a small portion of the training data is sufficient to adopt a QR-based model trained with non-conversational samples to work well on CANARD.

\begin{table}[t]
	\centering
	\caption{Extractive QA results on the CANARD test set. F1 and EM is calculated for both answerable and unanswerable questions, while NA Acc only for unanswerable questions.}
	\label{table:quac}
			\begin{tabular}{llccc}
				\toprule
				QA Input & Training Set & EM & F1 & NA Acc \\
				\midrule
				Original & CANARD-O & 38.68 & 53.65 & 66.55 \\
				Original + 1-DT &  & 42.04 & 56.40 & 66.72 \\
				Original + 2-DT &  & 41.29 & 56.68 & 68.11 \\
				Original + 3-DT &  & 42.16 & 56.20 & 68.72 \\
				\midrule
				Original & CANARD-H & 39.44 & 54.02 & 65.42 \\
				Human &  & 42.36 & 57.12 & 68.20 \\
				\midrule
				Original & MultiQA $\rightarrow$ & 41.32 & 54.97 & 65.84 \\
				Original + 1-DT & CANARD-H  & 43.15 & 57.03 & 68.64 \\
				Original + 2-DT &  & 42.20 & 57.33 & 69.42 \\
				Original + 3-DT &  & 43.29 & 57.87 & \textbf{71.50} \\
				Co-reference &  & 42.70 & 57.59 & 66.20 \\
				PointerGenerator &  & 41.93 & 57.37 & 63.16 \\
				CopyTransformer &  & 42.67 & 57.62 & 68.02 \\
				Transformer++ &  & \textbf{43.39} & \textbf{58.16} & 68.29\\
				Human &  & 45.40 & 60.48 & 70.55\\
				\bottomrule
			\end{tabular}
\end{table}

\paragraph{\textbf{RQ3}: What is the proportion of errors contributed by each of the components (traceability)?}
We measure the effect from question rewriting for each of the questions by comparing the answers produced for the original, the model-rewritten (Transformer++) and the human-rewritten question (see Tables~\ref{table:trec_breakdown}-\ref{table:quac_breakdown}).
This approach allows us to pinpoint the cases, in which QR contributes to the QA performance, and distinguish them from cases in which the answer can be found using the original question as well.

Assuming that humans always produce correct question rewrites, we can attribute all cases in which these rewrites did not result in a correct answer as errors of the QA component (rows 1-4 in Tables~\ref{table:trec_breakdown}-\ref{table:quac_breakdown}).
The next two rows 5-6 show the cases, where human rewrites succeeded but the model rewrites failed, which we consider to be a likely error of the QR component.
The last two rows are true positives for our model, where the last row combines cases where the original question was just copied without rewriting (numbers in brackets) and other cases when rewriting was not required.
Since there is no single binary measure for the answer correctness, we select different cut-off thresholds for our QA metrics.

The majority of errors stem from the QA model: 29\% of the test samples for retrieval and 55\% for extractive estimated for P@1 and F1, comparing to 11\% and 5\% for QR respectively.
Note that it is a rough estimate since we cannot automatically distinguish the cases that failed both QA and QR.

Overall, we observe that the majority of questions in extractive QA setup can be correctly answered without rewriting or accessing the conversation history.
In other words, the extractive QA model tends to return an answer even when given an incomplete ambiguous question.
This finding also explains the low NA Acc results reported in Table~\ref{table:quac}.
Our results provide evidence of the deficiency of the reading comprehension setup, which was also reported in the previous studies on non-conversational datasets~\cite{DBLP:conf/emnlp/JiaL17,DBLP:conf/iclr/LewisF19,DBLP:conf/acl/SinghGR18}.

In contrast, in the retrieval QA setup, only 10\% of the questions in TREC CAsT were rewriten by human annotators that did not need rewriting to retrieve the correct answer.
These results highlight the difference between the retrieval and extractive QA training/evaluation setup.
More details on our error analysis approach and results can be found in \citet{DBLP:journals/corr/abs-2010-06835}.



\begin{table}[t]
    	\caption{Break-down analysis of all retrieval QA results for the TREC CAsT dataset. Each row represents a group of QA samples that exhibit similar behaviour. $\checkmark$ indicates that the answer produced by the QA model was correct or $\times$ -- incorrect, according to the thresholds provided in the right columns. We consider three types of input for every QA sample: the question from the test set (Original), generated by the best QR model (Transformer++) or rewritten manually (Human). The numbers correspond to the count of QA samples for each of the groups. The numbers in parenthesis indicate how many questions do not require rewriting, i.e., should be copied from the original.}
	\centering
	\label{table:trec_breakdown}
			\begin{tabular}{ccccccc}
				\toprule
				 & & & P@1  & \multicolumn{3}{c}{NDCG@3} \\
				Original & QR & Human & = 1 & > 0 & $\geq$ 0.5 & = 1 \\
				\midrule
				$\times$ & $\times$ & $\times$ & 49 (14) & 10 (1) & 55 (20) & 154 (49) \\ 
				$\checkmark$ & $\times$ & $\times$ & 0 & 0 & 0 & 0 \\
				$\times$ & $\checkmark$ & $\times$ & 2 & 0 & 1 & 0 \\
				$\checkmark$ & $\checkmark$ & $\times$ & 0 & 1 & 1 & 0 \\
				\midrule
				$\times$ & $\times$ & $\checkmark$ & 19 & 10 & 25 & 4 \\
				$\checkmark$ & $\times$ & $\checkmark$ & 0 & 1 & 0 & 0 \\
				$\times$ & $\checkmark$ & $\checkmark$ & 48 & 63 & 47 & 11 \\
				$\checkmark$ & $\checkmark$ & $\checkmark$ & 55 (37) & 88 (52) & 44 (33) & 4 (4)  \\ 
				\midrule
				 \multicolumn{3}{c}{} & Total & \multicolumn{3}{r}{173 (53)} \\
				\bottomrule
			\end{tabular}
\end{table}

\begin{table}[t]
\caption{Break-down analysis of all extractive QA results for the CANARD dataset, similar to Table~\ref{table:trec_breakdown}.}
	\centering
	\label{table:quac_breakdown}
			\begin{tabular}{cccccc}
				\toprule
				Original & QR & Human & F1 > 0 & F1 $\ge$ 0.5 & F1 = 1  \\
				\midrule
				$\times$ & $\times$ & $\times$ & 847 (136) & 1855 (235) & 2701 (332) \\
				$\checkmark$ & $\times$ & $\times$ & 174 & 193 & 181 \\
				$\times$ & $\checkmark$ & $\times$ & 19 & 35 (2) & 40 (1) \\
				$\checkmark$ & $\checkmark$ & $\times$ & 135 & 153 & 120 \\
				\midrule
				$\times$ & $\times$ & $\checkmark$ & 141 & 288 & 232 \\
				$\checkmark$ & $\times$ & $\checkmark$ & 65 (1) & 57 (1) & 40 \\
				$\times$ & $\checkmark$ & $\checkmark$ & 226 & 324 & 269 \\
				$\checkmark$ & $\checkmark$ & $\checkmark$ & 3964 (529) & 2666 (428) & 1988 (333) \\
				\midrule
				 \multicolumn{3}{c}{} & Total & \multicolumn{2}{r}{5571 (666)} \\
				\bottomrule
			\end{tabular}
\end{table}

\section{Conclusion}
\label{sec:conclusion}



We showed in an end-to-end evaluation that question rewriting is effective in extending standard QA approaches to a conversational setting.
Our results set the new state-of-the-art on the TREC CAsT 2019 dataset.
The same QR model also shows superior performance on the answer span extraction task evaluated on the CANARD/QuAC dataset. 
Based on the results of our analysis, we conclude that QR is a challenging but also very promising task that can be effectively implemented into conversational QA approaches.

We also confirmed that the QR metric provides a good indicator for the end-to-end QA performance.
Thereby, it is reliable to be used for the QR model selection, which would help to avoid more costly end-to-end evaluation in the future. 

The experimental evaluation and the detailed analysis we provide increases our understanding of the main error sources.
QR performance is sufficiently high on both CANARD and TREC CAsT datasets, while the QA performance even given human rewritten questions for both tasks lags behind.
This result suggest that the major improvement in conversational QA will come from improving the standard QA models.


In future work we would like to evaluate QR performance with the joint model integrating passage retrieval and answer span extraction.
Recent results reported by \citet{DBLP:conf/sigir/Qu0CQCI20} indicate that there is sufficient room for improvement in the history modeling phase.

On the other hand, the QR-QA type of architecture is generic enough to incorporate other types of context, such as a user model or an environmental context obtained from multi-modal data (deictic reference).
Experimental evaluation of QR-QA performance augmented with such auxiliary inputs is a promising direction for future work.

\section*{Acknowledgements}
We would like to thank our colleagues Srinivas Chappidi, Bjorn Hoffmeister, Stephan Peitz, Russ Webb, Drew Frank, and Chris DuBois for their insightful comments.

\bibliographystyle{ACM-Reference-Format}
\bibliography{refs}

\end{document}